\documentclass[preprint, 3p, twocolumn]{elsarticle}
\usepackage{graphicx}
\usepackage{booktabs}
\usepackage{tabularx}
\usepackage[normalem]{ulem}
 \useunder{\uline}{\ul}{}
 \usepackage{rotating}
 \usepackage{dsfont}
 \usepackage{textcomp}
\usepackage{gensymb}
\usepackage{color}
\def\Manu#1{\textcolor{black}{#1}}
\newcommand{\bit}[1]{\textit{\textbf{#1}}}
\usepackage{graphicx,xcolor}
\usepackage{epstopdf}
\usepackage{setspace}
\usepackage[T1]{fontenc}
\usepackage{amssymb,amsfonts,amsmath}
\usepackage{setspace}
\usepackage{subcaption}
\usepackage{mathtools}
\usepackage{etoolbox}
\biboptions{sort&compress}
\newcolumntype{Y}{>{\centering\arraybackslash}X}
\usepackage[switch, mathlines]{lineno}
\usepackage{upgreek}

\let\oldequation\equation
\let\oldendequation\endequation

\renewenvironment{equation}
  {\linenomathNonumbers\oldequation}
  {\oldendequation\endlinenomath}

\makeatletter
\def\ps@pprintTitle{%
 \let\@oddhead\@empty
 \let\@evenhead\@empty
 \def\@oddfoot{\footnotesize\itshape
       Preprint submitted to \ifx\@journal\@empty Royal Society Interface
       \else\@journal\fi\hfill\today}
 \let\@evenfoot\@oddfoot}
\makeatother

\begin{document}

\title{\Manu{Surface Energy and Separation Mechanics of Droplet Interface Phospholipid Bilayers}} 
\author[yogi,equal]{Y. Huang}
\author[yogi,equal]{V. Chandran Suja}
\author[yogi,javi]{J. Tajuelo}
\author[yogi]{G.G. Fuller}
\ead{ggf@stanford.com}
\address[yogi]{Department of Chemical Engineering, Stanford University, Stanford, California 94305}
\address[javi]{Departamento de Física Interdisciplinar, Universidad Nacional de Eduación a Distancia UNED, Madrid 28040, Spain}
\address[equal]{Equal contribution}
\begin{abstract}
 Droplet interface bilayers are a convenient model system to study the physio-chemical properties of phospholipid bilayers, the major component of the cell membrane. The mechanical response of these bilayers to various external mechanical stimuli is an active area of research due to implications for cellular viability and development of artificial cells. In this manuscript we characterize the separation mechanics of droplet interface bilayers under step strain using a combination of experiments and numerical modeling. Initially, we show that the bilayer surface energy can be obtained using principles of energy conservation. Subsequently, we subject the system to a step strain by separating the drops in a step wise manner, and track the evolution of the bilayer contact angle and radius. The relaxation time of the bilayer contact angle and  radius, along with the decay magnitude of the bilayer radius were observed to increase with each separation step. By analyzing the forces acting on the bilayer and the rate of separation, we show that the bilayer separates primarily through the peeling process with the dominant resistance to separation coming from viscous dissipation associated with corner flows.  Finally, we explain the intrinsic features of the observed bilayer separation by means of a mathematical model comprising of the Young-Laplace equation and an evolution equation. We believe that the reported experimental and numerical results extend the scientific understanding of lipid bilayer mechanics, and that the developed experimental and numerical tools offer a convenient platform to study the mechanics of other types of bilayers.
\end{abstract}

\maketitle
\section{Introduction}
The phospholipid bilayer is an important component of cell membrane that regulates material transport in a wide range of conditions \cite{lombard2014once,naumann2008protein,bello2017lipid}. It was first extracted and identified by Gorter and Grendel in 1925 \cite{gorter1925bimolecular}. Subsequent research, notably by Singer and Nicolson through their fluid mosaic model of the cell membrane, improved the physio-chemical understanding of the lipid bilayer \cite{singer1972fluid}. 
Currently, lipid bilayers are commonly used to study the ion transport and protein interaction across cell membranes \cite{gambale1982properties, ter1993interaction,villar2011formation}. One important property of the bilayer that influences the above transport processes as well as other cellular functions such as membrane fusion, ion binding and integral protein activity, is the bilayer surface energy \cite{barlow2018measuring,ohki1985divalent,lee2004lipids}. Another important aspect of bilayers that influences cellular functioning are the interaction forces within the bilayer \cite{lee2018revisiting,orozco2013interaction}, which is explored in the present work through the separation of the two monolayers forming the bilayer. Convenient investigation of the above bilayer features requires in vitro lipid bilayer models. 

\begin{figure*}[!ht]
\centering
\includegraphics[width=\linewidth]{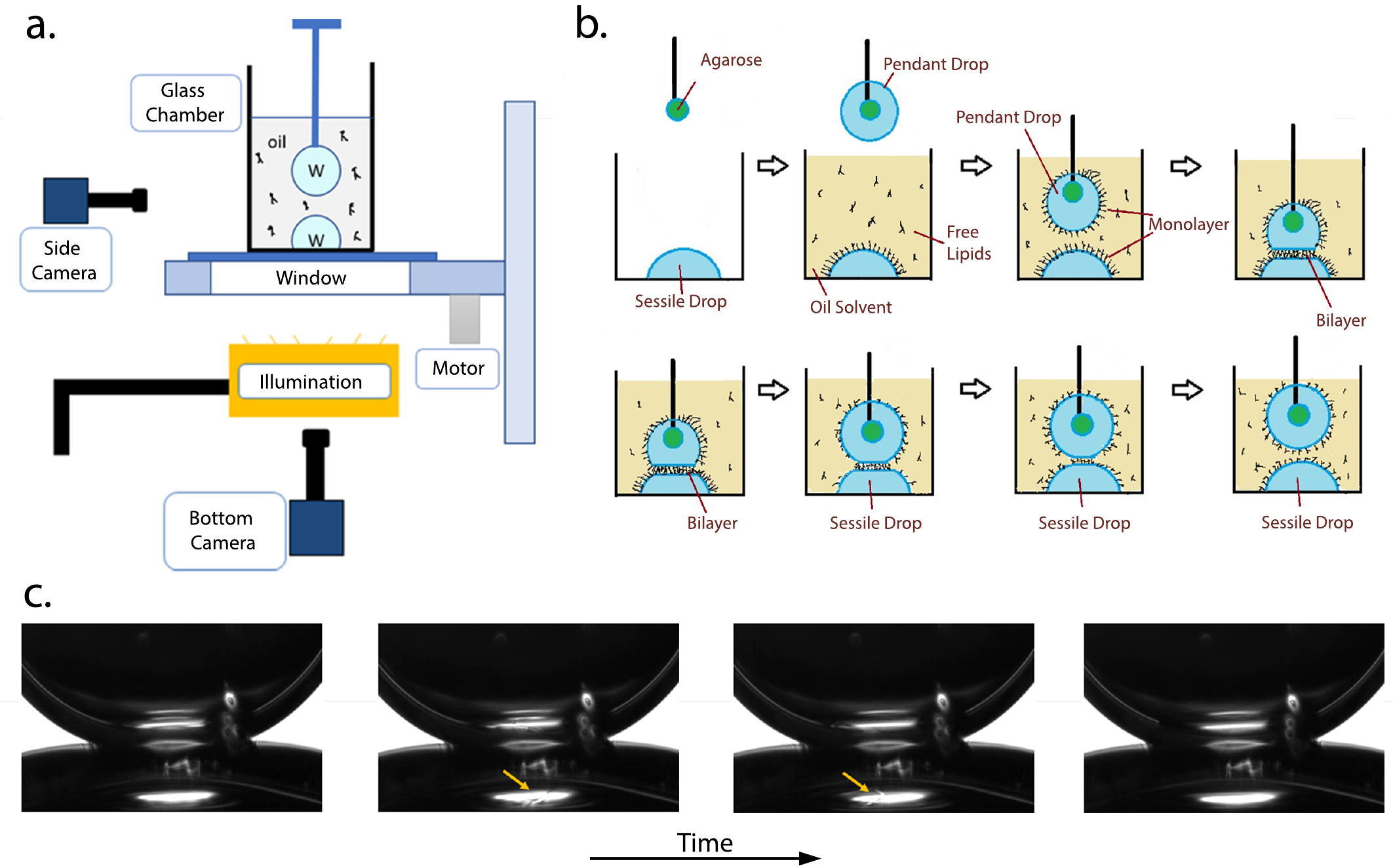} \caption{A schematic of the experimental setup and protocol. ({\bf a.}) Schematic of the Interfacial Drainage Dilatational and Stability Stage (I-DDiaSS) setup along with the labeled components. ({\bf b.}) The experimental protocol followed for the reported experiments: a sessile droplet of 1M KCl solution is pipetted on the bottom of a glass chamber. A second droplet of the same solution is pipetted onto the tip of a capillary, where a small volume of agarose gel was previously attached by immersing the capillary into a heated 3\% (wt/vol) agarose gel solution. The glass chamber is filled with 0.2\,mL of a 10\,mM solution of DPhPC in hexadecane, and the pendant droplet is submerged. The two droplets are kept apart for 15\,min to allow for the formation of a DPhPC monolayer on the droplets interface. Then, the droplets are brought into close contact. After a few minutes ($\sim5\mbox{\,min}$), the drainage of the oil thin film is complete and the bilayer is formed. The bilayer formation is visible as i) a front crossing the droplets contact area in the side camera, and ii) the apparition of a boundary in the bottom camera. Then, the droplets are separated in a step wise manner while the side images and the vertical stage displacement are recorded until the bilayer is finally separated. ({\bf c.}) A sequence of images obtained from the side camera showing the formation of the bilayer. The yellow arrows indicate the advancing front of the bilayer as it forms between the two droplets.} \label{fig:ExperimentalSetup}
\end{figure*}

Several techniques have been reported in literature for fabricating in vitro bilayers. These include solid supported lipid bilayers (SLBs), giant uni-lamellar vesicles (GUVs), black lipid membranes (BLMs) and droplet interface bilayers (DIB) \cite{beltramo2016millimeter,funakoshi2006lipid,evans1987physical}. The droplet interface bilayer is an attractive technique for creating artificial bilayers and has been widely used in the characterization of multiple bilayer features \cite{bayley2008droplet, thiam2012stability}. One key advantage of this technique, as compared to SLBs and BLMs, is that droplet interface bilayers can be conveniently imaged using cameras. The captured drop shapes, prior to the formation of the bilayer, can be used to calculate the monolayer surface tension employing the Young-Laplace equation \cite{alvarez2009non, neeson2014compound}. After the formation of the bilayer, the captured drop profiles conveniently allow us to track the mechanical response of the drops and the bilayer to external forcing. Further, droplet interface bilayers can be easily integrated with electrodes for measuring electrical properties \cite{najem2017mechanics} and with arrangements such as AFM cantilever beams for measuring forces within the bilayer \cite{frostad2014direct}. Because of these advantages, droplet interface bilayers have been commonly used to study the electrical properties \cite{hwang2007electrical,punnamaraju2011voltage}, permeability \cite{lee2018static, fleury2020enhanced}, bilayer surface energy \cite{dixit2012droplet,taylor2015direct}, interaction and adhesion forces \cite{frostad2014direct,venkatesan2015adsorption,yanagisawa2013adhesive}, and bilayer separation mechanics in a variety of conditions \cite{chatkaew2009dynamics,freeman2016mechanoelectrical,najem2017mechanics}.

Despite the vast literature on the mechanics of droplet interface bilayers and vesicle separation, the separation mechanics (unzipping) of a single bilayer in response to a step strain deformation has not been previously considered. We address this knowledge gap in this manuscript through a combination of experiments and mathematical modeling. The experiments are performed on a custom built setup called as the interfacial drainage dilatational and stability stage (I-DDiaSS), where the relative positions of a sessile and pendant drop can be precisely manipulated. The experiments are supported by a mathematical model, which serves to physically explain the observed bilayer separation mechanics. The rest of the manuscript is organized as follows. In Section \ref{sec:methods} we describe in detail the experimental setup, methodology, and the mathematical model. The key findings from this study are reported in Section \ref{sec:results}, where we show (i) a new method to calculate the bilayer surface energy employing principles of energy conservation, (ii) experimental data highlighting the intrinsic features of droplet interface bilayer separation under step strain, and (iii) a quantitative comparison of the experimental data against the predictions from the mathematical model. Finally we conclude the manuscript by discussing key areas for future research.

\section{Materials and Methods}\label{sec:methods}
\subsection{Materials}
DPhPC (1,2-diphytanoyl-sn-glycero-3-phosphocholine) was used as the model lipid for generating the phospholipid bilayers reported in this manuscript. Prior to the start of the experiments, DPhPC (Catlog no: 850356; Avanti Polar Lipids Inc., Alabaster, Alabama) was extracted from the suspending chloroform solution in two steps. Initially, the bulk of the chloroform was evaporated off by gently blowing a stream of nitrogen for $45$ minutes. Subsequently, the residual lipid film was vacuum dried for another $60$ minutes. The chloroform free lipids were then dissolved in hexadecane to give a final concentration of 10 mM.   

Agarose gel used as a core to support the pendent drop (Fig.\ref{fig:ExperimentalSetup}), was created using agarose powder purchased from Thermo Fisher Scientific (Catlog no: BP164100). $300$ mg of the powder was mixed with $10$ ml of distilled water at high temperature, and then cooled down to make the agarose gel \cite{holden2007functional,leptihn2013constructing}. 1M KCl solution was used for preparing the aqueous sessile and pendant droplets.

\subsection{Experimental Setup}
To create and separate the droplet interface phospholipid bilayers, we use the Interfacial Drainage Dilatational and Stability Stage (I-DDiaSS) setup \cite{poulos2010automatable,bochner2017droplet}. As shown in Fig.\ref{fig:ExperimentalSetup}a, I-DDiaSS consists of a glass chamber to hold the hexadecane solution and a sessile drop. The sessile drop is pinned at the center of the glass chamber bottom using a circular scratch (not visible in the figure). The glass chamber sits atop a motorized stage (Newport TRA12PPD and SMC100PP), which enables us to move the sessile drop relative to a pendant drop. The pendant drop is held in place by a blunt capillary needle (ID: 0.58 mm OD: 0.81 mm) having agarose gel at its tip. The agarose gel in the core of the pendant drop anchors the drop\cite{leptihn2013constructing} and prevents dripping.  The drop profiles and the bilayer radius are obtained through a side camera (IDS UI 3060CP) using principles of shadowgraphy. The formation of the bilayer (\bit{see Supplementary Materials}) is also confirmed via a bottom camera (Lumenera Infinity3-3UR).    


\subsection{Experimental protocol}
At the start of an experiment, we pipetted a predetermined amount of KCl solution onto the circular scratch on the bottom of the chamber to form a sessile drop, and 0.2 mL of the hexadecane solution with the DphPC lipid was gently added into the chamber until it fully covered the sessile drop. Then, a capillary with agarose was placed above the sessile drop, and another drop of the KCl solution was pipetted onto the agarose to form a pendant drop of volume $V_p = 1$ $\upmu$L. Then, two precision stages that control the position of the glass chamber along the horizontal axes are used to place the two droplets along the same vertical axis. To study the effects of drop size on the mechanics of droplet interface bilayers, we varied the sessile drop volume and obtained three different sessile to pendant drop volume ratios, namely $V_s/V_p=0.5, \; 1, \;1.5$.    


To form the bilayers, the motorized stage translated the pendant drop into the lipid oil phase. Both the pendant and sessile drops were then aged for 15 minutes in order to allow for the formation of the lipid monolayers at the oil-water interfaces \cite{leptihn2013constructing}. After aging, the sessile drop was slowly pushed against the pendant drop for approximately 0.35 mm and then held in place. The thin liquid film between the pendant and sessile droplets drains until the lipid monolayers are close enough to form a bilayer. As the bilayer forms, we observe an advancing front that crosses the entire contact area between the droplets in a short time (see  Fig.\ref{fig:ExperimentalSetup}c). This important event confirms the formation of a bilayer between the droplets. Further details confirming the formation of the bilayer are available in the \bit{Supplementary Materials}   For all the reported experiments, the initial bilayer radius was 0.25 mm. 

To conduct the bilayer separation experiments, we again used the motorized stage to pull the sessile drop away from the pendant drop in a step-wise manner at a velocity of 0.05 mm/s for one second. The step size ($d$) has a constant value of 0.05 mm, resulting in step strain of $d/R_a = 0.067$, where $R_a = 0.75$ is the apex drop curvature of the pendant drop. After each separation step we allowed the bilayer to relax for 2 minutes. This process was continued until the sessile and pendant drops separated completely. The entire process of the bilayer formation and separation was captured using both the side and bottom camera, and was subsequently analyzed utilizing MATLAB. All experiments mentioned in this paper were performed at room temperature. A schematic diagram of the above mentioned bilayer formation and separation on the I-DDiaSS is shown in Fig.\ref{fig:ExperimentalSetup}b.

\begin{figure}
\centering
    \includegraphics[width=\linewidth]{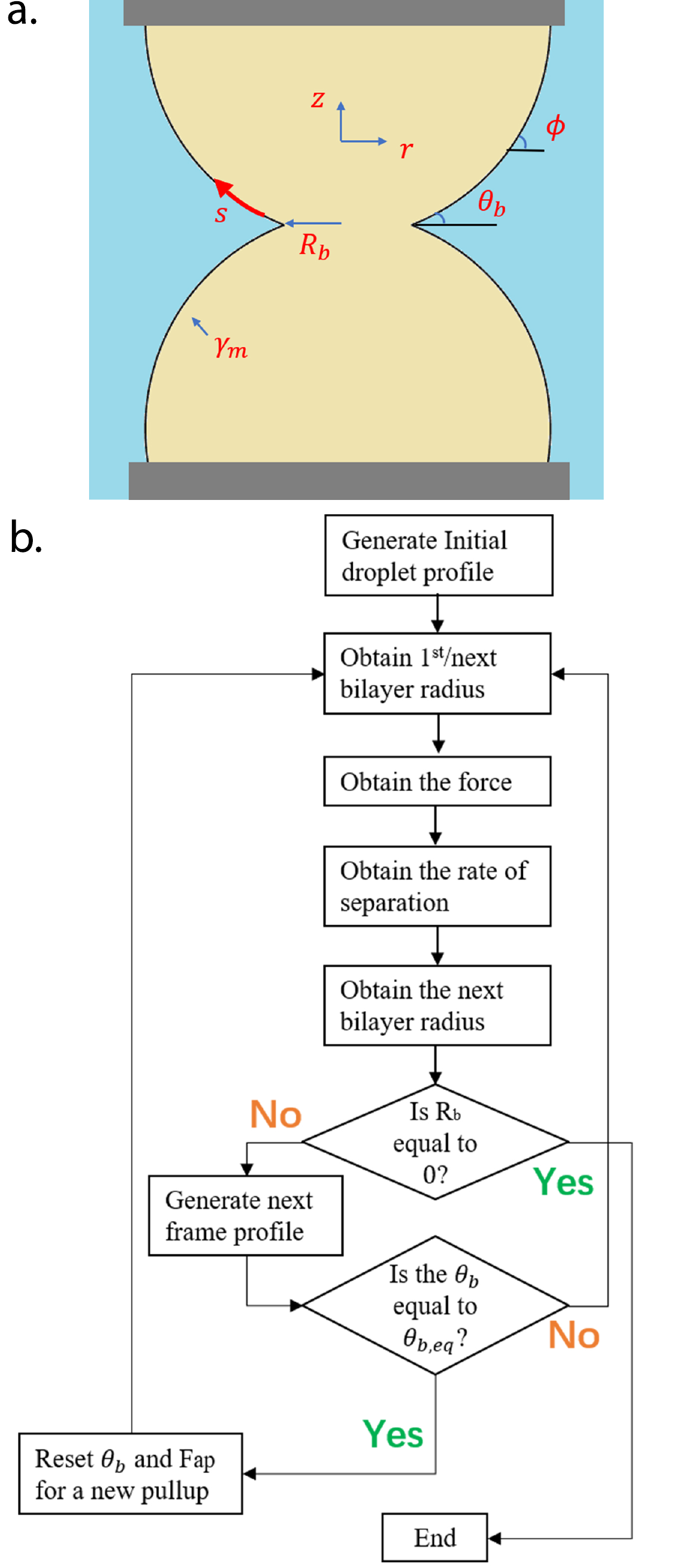} \caption{{\bf a.} A schematic of the pendant-sessile drop system used in the mathematical model. {\bf b.} Flow chart showing the iterative solution of the equations governing the bilayer separation.} \label{fig:SimNotation}
\end{figure} 

\subsection{Mathematical model for droplet interface bilayer separation}\label{subsec:Model}
Here we will develop a mathematical model based on the Young-Laplace equation for capturing the separation mechanics of droplet interface bilayers. In this simplified axisymmetric model, as shown in Fig.\ref{fig:SimNotation}a, we ignore the agarose attached to the needle. We further assume the drop profiles to be symmetric with respect to the plane of the bilayer. Physically, this assumption holds at low Bond number for sessile and pendant drops of comparable sizes. Under these assumptions, the non-dimensional Young-Laplace equation governing the shapes of the drops can be written as,  
\begin{align}
    \frac{d\phi}{d\bar{s}}&=2-\text{Bo}\bar{z}-\frac{\sin{\phi}}{\bar{r}}\\
    \frac{d\bar{r}}{d\bar{s}}&=\cos{\phi}\\
    \frac{d\bar{z}}{d\bar{s}}&=\sin{\phi}
\end{align}

\noindent where $\bar{s}$ is the arc length along the pendant drop, measured from the edge of the bilayer and non-dimensionalized by the apex drop curvature ($R_a$). $\phi$ is the angle between the tangent to the pendant drop profile and the horizontal. $\bar{r}$ and $\bar{z}$ represent the non-dimensional cylindrical coordinates of the bilayer interface. Bo is the Bond number denoted by $\text{Bo}=\Delta \rho g R_a^2/\gamma_m$, where $\gamma_m$ is the monolayer surface tension, and  $\Delta \rho$ is the density difference between aqueous and oil phases. For low Bond number drops having similar sizes, the bilayer radius is set by the following force balance in the vertical direction (see \bit{Supplementary Materials} for details),  

\begin{equation}
   -\bar{R}_b^2 +\bar{F}_{ap} +  \bar{R}_b\sin{\theta_b} =0
\end{equation}
where $\bar{R}_b$ is the normalized bilayer radius, $\bar{F}_{ap}$ is the external force acting on the drop non-dimensionalized by $2\pi R_a \gamma_m$, $\gamma_m$ is the monolayer surface tension and $\theta_b$ is the contact angle between the monolayer and bilayer as illustrated in Fig.\ref{fig:SimNotation}a.  Physically, the first term is the non-dimensional Laplace pressure originating from the deformation of the drops along the plane of the bilayer, the second term is the external force pushing the drops against each other and the third term is the non-dimensional interfacial tension acting along the periphery of the bilayer. 


\begin{figure*}[!h]
\centering
\includegraphics[width=\linewidth]{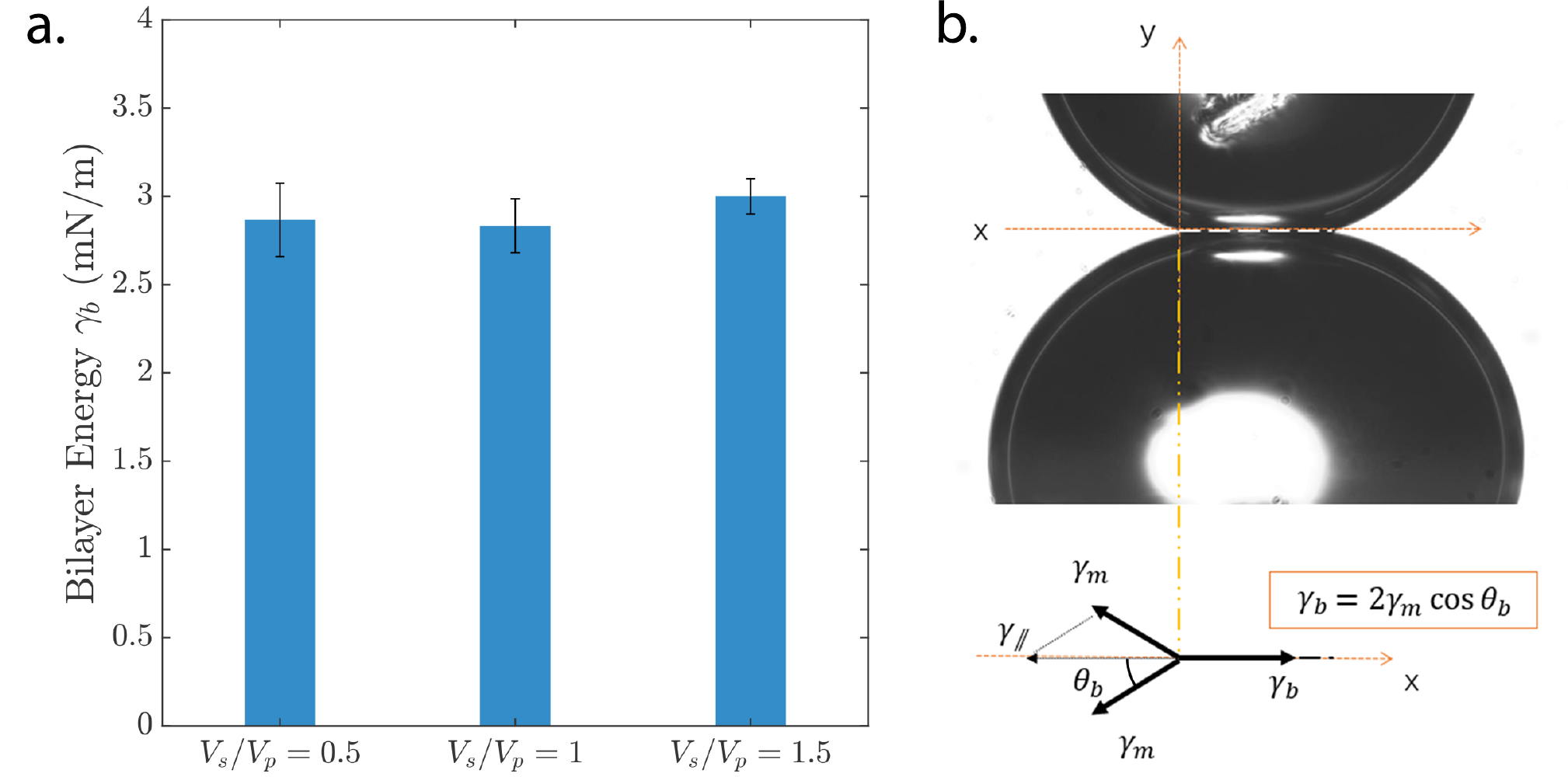} \caption{Bilayer surface energy. {\bf a.} Bilayer energy computed from Eq.\ref{eq:EnergyRegression} as a function of the ratio of $V_s/V_p$. Here  $V_s$ is the sessile drop volume and $V_p$ is the pendant drop volume. Experimental error bars are included. {\bf b.} Illustration of bilayer surface energy calculation using the Neumann force balance.} \label{fig:BilayerEnergy}
\end{figure*}

Finally we relate the excess vertical interfacial tension acting on the bilayer, $\gamma_\perp - \gamma_{\perp,eq}$, to the bilayer radius $R_b$ and the separation velocity of the bilayer, $v = -\frac{dR_b}{dt}$, where $\gamma_\perp =\gamma_m \sin \theta_b$ and $\gamma_{\perp,eq}$ is the value of $\gamma_{\perp}$ when $v = 0$. Expressing $\gamma_\perp$ in terms of the variables in Fig.\ref{fig:SimNotation}a closes the system of equations and gives the following non-dimensional evolution equation for $v$,


\begin{equation}\label{eq:SeperationVelocity}
    \bar{v}  = \frac{\bar{b}(\sin \theta_b - \sin \theta_{b,eq})}{K \bar{R_b}},
\end{equation}
where, $\bar{v} = \frac{\mu v}{\gamma_m}$ is the non-dimensional separation velocity, $\mu$ the dynamic viscosity of the ambient oil phase,  $\theta_{b,eq}$ is the value of $\theta_b$ when the bilayer is in equilibrium, $\bar{b}$ the non-dimensional thickness of the bilayer and $K$ is a prefactor in the relationship between the excess force acting on the bilayer and the peeling rate of the bilayer (see \bit{Supplementary Materials} and \bit{Section \ref{subsec:ForceRate}} for more details).

The above system of equations is solved in an iterative process as illustrated in the flow chart in Fig \ref{fig:SimNotation}b. In the beginning, the initial droplet profile is obtained by solving the Young-Laplace equation with the relevant values of Bo, $R_b$ and the contact angle $\theta_b$. From this configuration, the contact angle and bilayer radius are extracted to give the corresponding $\gamma_\perp$ and $v$. The rate of separation then allows one to obtain the bilayer radius in the upcoming step, which is then used to generate a new droplet profile. The iterations proceed in steps of $0.05$ s for a total duration of $120\;s$, which matches the duration of a single step separation in our experiments. During this time, bilayer continuously evolves until it reaches the equilibrium state where $\theta_b$ equals $\theta_{b,eq}$. To initiate a new step, $\theta_b$ is altered to mimic the contact angle change following a pull-up in the actual experiment. $\bar{F}_{ap}$ is recalculated, and the above algorithm is repeated until the bilayer radius tends to zero, reflecting a complete separation of the bilayer. Further details regarding the model and its solution are available in the \bit{Supplementary Materials}.

\section{Results and Discussion}\label{sec:results}

\subsection{Bilayer Surface Energy}\label{subsec:bilayerenergy}
Bilayer surface energy is an important property that dictates the mechanics of bilayer separation by directly influencing the equilibrium bilayer contact angle. Here we calculate the surface energy of our DPhPC droplet interface bilayers using  two independent techniques.

Firstly, we can exploit principles of energy conservation to calculate bilayer surface energy. The total energy of the system ($E_{total}$) at equilibrium can be written as,   

\begin{equation}\label{eq:totEnergy}
    E_{total}  =  E_p + E_s^m + E_s^b
\end{equation}
where $E_p$ is the potential energy, $E_S^m$ is monolayer surface energy and $E_S^b$ is bilayer surface energy as a function of time. We can relate each of these energies to measurable physical quantities as follows, 
\begin{align}\label{eq:potenergy}
    E_p  &=\Delta \rho g V h\\ \label{eq:monenergy}
    E_s^m  &=\gamma_m S^m\\\label{eq:bienergy}
    E_s^b  &=\gamma_b S^b
\end{align}
 Here $\gamma_b$ is the bilayer surface tension, and $\gamma_m$ is monolayer surface tension, which we found to be equal to 1.62 mN/m using the pendant drop technique. $s^m$ and $s^b$ are monolayer and bilayer area, $V$ is the combined volume of both the drops and $h$ is the vertical co-ordinate of the center of mass of the system. Substituting Eqns. \ref{eq:potenergy}, \ref{eq:monenergy} and \ref{eq:bienergy} in to Eq.\ref{eq:totEnergy}, and rearranging, we obtain,
\begin{equation}\label{eq:EnergyRegression}
    \Delta \rho g V h+ \gamma_m s^m =  - \gamma_b s^b + E_{total}.
\end{equation}
The above equation produces a linear relaxation between $\Delta \rho g V h+ \gamma_m s^m$ and $s^b$. By measuring and plotting the LHS of Eq.\ref{eq:EnergyRegression} as a function of $s_b$,(\textbf{see Supplementary Materials}) we can obtain $\gamma_b$ from the slope of the best fit line to the data.  $\gamma_b$ obtained from the energy analysis is shown in Fig.\ref{fig:BilayerEnergy}a for three different drop size ratios. As expected, we find that the bilayer energy calculated from the above technique is independent of the size of the drops. Averaging across all measurements, we find $\gamma_b = 2.90 \pm 0.16$ mN/m.  

Secondly, a simple force balance along the contact line can also be used to calculate the bilayer surface energy \cite{barlow2018measuring}. Resolving the surface tensions along the tangent to the bilayer (Fig.\ref{fig:BilayerEnergy}b), we obtain the following expression for the bilayer energy,
\begin{equation}\label{eq:bilayerForce}
    \gamma_b = 2\gamma_m \cos \theta_b,
\end{equation}

\noindent where $\theta_b$ is half the angle between pendant and the sessile drops at equilibrium. Consistent with previous reports \cite{venkatesan2015adsorption,najem2017mechanics}, we find $\theta_b \approx 20^o$ from our experiments. Substituting this value of $\theta_b$ in Eq.\ref{eq:bilayerForce}, we find $\gamma_b \approx 3$ mN/m. This value of $\gamma_b$ is very similar to that obtained from the energy analysis, indicating that the energy analysis is a comparable technique that can be used to calculate the bilayer surface tension. Since the energy analysis does not require the contact angle, which is often difficult to measure accurately, this method is a valuable alternative for bilayer surface tension calculations.    
\begin{figure*}
\centering
     \includegraphics[width=0.99\linewidth]{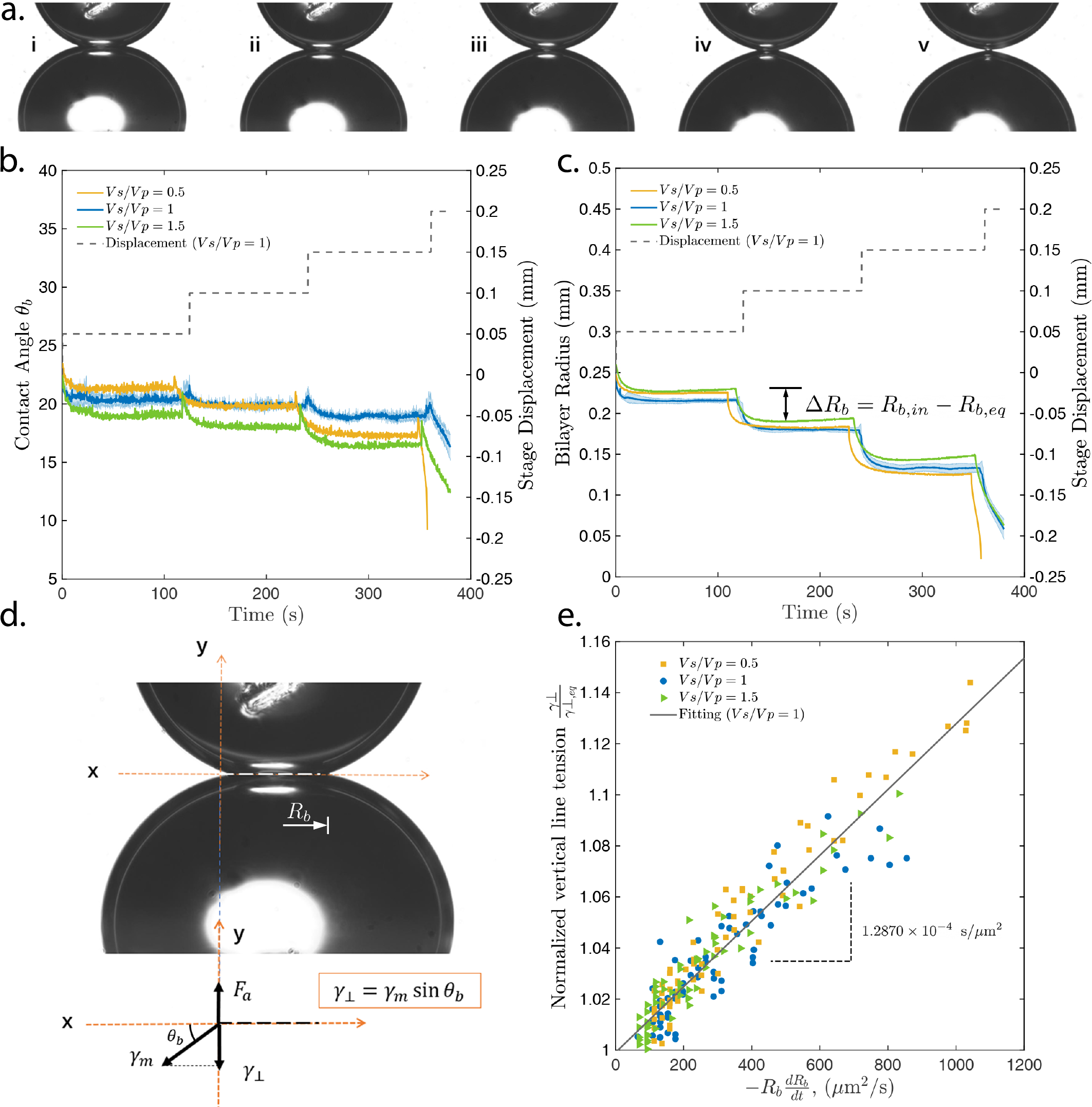} \caption{Mechanics of bilayer separation. {\bf a.} Sequence of images showing the bilayer at the beginning of each step and just before separation. {\bf b.} Evolution of the contact angle $\theta_b$ for different sessile to pendant drop volume ratios. The shaded error bar for the case where $V_s/V_p=1$ indicates the standard deviation obtained from three independent measurements. The dashed line shows the corresponding step displacements of the stage for pulling the sessile drop away from the pendant drop. The step size ($d$) has a constant value of 0.05 mm,  resulting in a step strain of $d/R_a = 0.067$, where $R_a$ is the apex drop curvature of the pendant drop.  {\bf c.} Evolution of the bilayer radius $R_b$. The shaded error bar for the case where $V_s/V_p=1$ indicates the standard deviation obtained from three independent measurements. The dashed line again shows the corresponding step displacements of the stage. $\Delta R_b$ is the bilayer radius decay and is calculated as the difference between the bilayer radius at the start of the step ($R_{b,in}$) and that at the plateau when a new equilibrium value is reached ($R_{b,eq}$). {\bf d.} A schematic illustrating the calculation of the vertical surface tension component using the Neumann force balance. The bilayer radius $R_b$ is also marked for clarity. {\bf e.} Normalized vertical component of the surface tension $\gamma_\perp$ as function of the product of the bilayer radius and the bilayer separation velocity $v= -\frac{dR_b}{dt}$. Here $\gamma_{\perp,eq}$  is the value of $\gamma_\perp$ at $\frac{dR_b}{dt}=0$. The solid line is a linear fit to the data and has a slope of $1.2870\times 10^{-4}\;\; \text{s}/\mu \text{m}^2$.} \label{fig:Mechanics}
\end{figure*}

\subsection{Mechanics of Bilayer Separation under Step Strain}
The separation mechanics of droplet interface bilayers have been previously investigated under constant droplet separation rates \cite{frostad2014direct} and under a constant force \cite{brochard2003unbinding}. Different from the separation of two adhered bilayers, here we report the separation mechanics (unzipping) of a single bilayer in response to a  step strain deformation. For this purpose, the droplets were separated in multiple steps by periodically displacing the stage supporting the sessile drop. During this process, we tracked the mechanical response of the bilayer to the applied step strain by measuring the contact angle $\theta_b$ and the bilayer radius $R_b$.   

\subsubsection{Contact Angle and Bilayer Radius}
The evolution of the bilayer contact angle and bilayer radius (see Fig.\ref{fig:Mechanics}d for the definition) in response to the applied step strain for three different droplet size ratios is shown in Fig.\ref{fig:Mechanics}b and Fig.\ref{fig:Mechanics}c respectively. The shaded error bar for the case where $V_s/V_p = 1$ indicates the standard deviation obtained from three independent measurements. The corresponding stage displacement profiles used in the experiment are shown using dashed lines.

These measurements reveal, at first sight, a couple of interesting features of bilayer separation mechanics under step strain. Firstly, following every step separation up to the last one, we observe a step increase of the contact angle, followed by a gradual relaxation of the contact angle to its equilibrium value. A caveat is in order here: notice in Fig. \ref{fig:Mechanics}b that, for $V_s/V_p\neq 1$, the equilibrium values of $\theta_b$ before and after each separation step are slightly different. This indicates that the lipid bilayer now is no longer flat, which could induce a measurement error as we track the evolution of the contact angle by looking at the drop profiles a few pixels away from the contact point. In any case, the fact that $\theta_b$ reaches a non time-dependent value after each separation step allows us to safely assume that the geometry is such that $\gamma_m$ and $\gamma_b$ are balanced. Regarding the bilayer radius, we also see that it decays to a new equilibrium value following every step until complete bilayer separation is achieved. Interestingly, the magnitude of the bilayer radius decay, which we will refer to as $\Delta R_b$, increases with each step. This is a characteristic of bilayer separation between curved surfaces and will be explained in more detail in \bit{Section \ref{subsec:Simulation}}. Secondly, we observe that at the last step, there is a rapid change in the angle and radius close to the complete separation of the bilayer. It is also worth noting that in the last step, the bilayers separated completely in all our experiments without the formation of thin strands of liquid, more commonly referred to as tethers in the literature \cite{frostad2014direct,waugh1982surface}.

All the above observations support that the mechanics of the separation of the droplet interface bilayers in our experiments are primarily a result of the peeling process. During the peeling process, the thickness of the bilayer remains a constant as its radius decreases \cite{frostad2014direct}. This is clearly the case up to the last step. Even in the last step, peeling is dominant as seen by the smooth change in the bilayer radius over time. The pulling process, where the bilayer thickness changes with little change in bilayer radius, plays a minor role in the reported bilayer separation and possibly occurs only in the last few seconds prior to complete separation.  In fact, in \bit{Section \ref{subsec:Simulation}} we will show that the observed variation in the bilayer radius (and contact angle) can be completely captured with a model that only considers peeling. Before we do so, let us examine the dynamics of bilayer separation by investigating the forces acting on the bilayer and the rates of separation.

\subsubsection{Force and Rate Analysis}\label{subsec:ForceRate}
Recall that in \bit{Section \ref{subsec:bilayerenergy}} we mentioned that a simple force balance along the contact line is made to obtain the bilayer surface energy. Similarly, we can perform a force balance normal to the contact line to obtain the vertical component of $\gamma_m$ acting on the bilayer perimeter. This quantity is denoted as $\gamma_\perp$, and is obtained as,
\begin{equation}
     \gamma_\perp = \gamma_m \sin \theta_b. \label{eq:VerticalForce}
\end{equation}

\noindent $\gamma_\perp$ drives the separation of the bilayer. When $\gamma_\perp$ exceeds the critical adhesive force per unit length, the bilayer starts to peel. The rate of peeling, which is denoted as $v$, can be obtained from the evolution of the bilayer radius as
\begin{equation}
    v = -\frac{dR_b}{dt}.
\end{equation}
The magnitude of $v$ naturally depends on the excess tension acting on the bilayer. To explicitly identify this correlation, we plot $\frac{\gamma_\perp}{\gamma_{\perp, eq}}$ as a function of $R_b v$ for three different droplet size ratios in Fig. \ref{fig:Mechanics}e. Here $\gamma_{\perp, eq}$  is the value of $\gamma_\perp$ at $v=0$, and physically represents the critical surface tension above which the bilayer starts to separate. Note the data in Fig. \ref{fig:Mechanics}e was obtained from the relaxation mechanics observed in a single step, specifically from the second step in Fig. \ref{fig:Mechanics}b and c. Plots of normalized surface tension versus $t$, $R_b v$ versus $t$ for the  first pull up, and the  lower and upper bound of the vertical forces  are available in the \bit{Supplementary Materials}.

\begin{figure*}[!h]
\centering
\includegraphics[width=\linewidth]{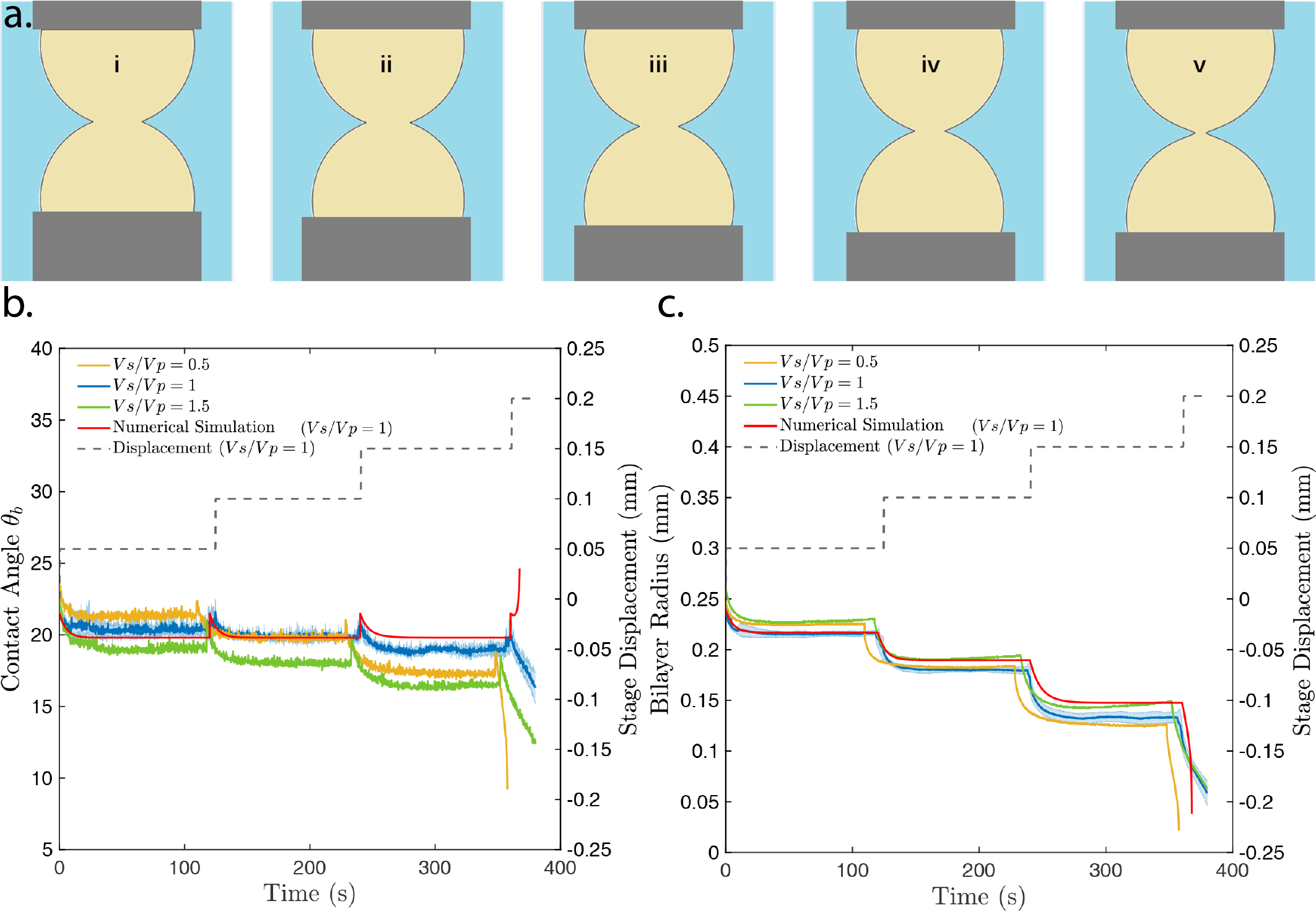} \caption{Results from the simulation {\bf a.} Simulated droplet profile at five important time points. \textbf{i-iv.} the first frame for the first, second, third and terminal step. \textbf{v.} Last frame for the terminal step, after that the droplets will be fully separated {\bf b.} Evolution of the contact angle $\theta_b$ by simulation marked by red line, compared with the previous experimental results {\bf c.} Evolution of the bilayer radius $R_b$ by simulation marked by red line, compared with the previous experimental results.} \label{fig:Simulation}
\end{figure*}

From Figure \ref{fig:Mechanics}e we see that $\frac{\gamma_\perp}{\gamma_{\perp, eq}}$ is a linear function of $R_b v$, with the slope of the best fit line remaining independent of the size ratios.  In agreement with prior studies \cite{frostad2014direct,brochard2003unbinding,chatkaew2009dynamics}, this correlation indicates that the peeling of non-specifically adhered phospholipid bilayers is retarded primarily by the viscous dissipation in the corners formed between the drops near the bilayer. Balancing the vertical surface tension with the viscous dissipation, we obtain the following expression \cite{chatkaew2009dynamics},   

\begin{equation}
    \frac{\gamma_\perp}{\gamma_{\perp,eq}} =1 -\left(\frac{\mu }{2b\gamma_{\perp,eq}}\right) R_b\frac{d R_b}{dt}
 \label{eq:JohnModel}
\end{equation}

\noindent where $\mu$ is the viscosity in the bulk oil phase and $b$ is the thickness of the lipid bilayer. Performing an order of magnitude analysis by taking $\mu$ to be of the order of 1 cP, $b$ to be of the order of the $1$ nm and $\gamma_{\perp,eq}$ to be of the order of $1$ mN/m, we see that term in the bracket of the above equation has a value of $5\times 10^{-4}\;\; \text{s}/\mu \text{m}^2$. This value is comparable to the experimentally determined slope of the linear fit between $\frac{\gamma_\perp}{\gamma_{\perp,eq}}$ and  $- R_b\frac{d R_b}{dt}$ (Fig.\ref{fig:Mechanics}e). This agreement confirms that viscous dissipation is the dominant mechanism retarding the separation of the tested phospholipid drops. Finally it is worth noting that this viscous resistance becomes negligible close to bilayer separation ($R_b \rightarrow 0$), and contributions from lubrication forces dominate as peeling gives way to the pulling mode of separation \cite{frostad2014direct}.

\subsection{Simulation}\label{subsec:Simulation}
Here we report the predictions from the mathematical model (see \bit{Section \ref{subsec:Model}}) obtained with the following values of the free variables - $\text{Bo}=0.1$, $\theta_{b,eq} = 19.8^o$ and $\gamma_m = 1.62$ mN/m. Except for the Bo, all the variables used in this model have values similar to those in the experiments, and Bo can be considered as a fitting parameter used to compare the simulations with the experiments. This is done in order to accurately recreate the drop profile in the absence of the agarose core supporting the pendant drop.


From solving the bilayer separation model, we obtain the evolution of the droplet profiles (Fig.\ref{fig:Simulation}a), the contact angle (Fig.\ref{fig:Simulation}b) and the bilayer radius (Fig.\ref{fig:Simulation}c). In Fig.\ref{fig:Simulation}a, we see the droplet profiles at the start of the four steps (i-iv) and right before separation (v). Qualitatively, we observe a good agreement between the simulated and experimentally obtained profiles (Fig. \ref{fig:Mechanics}a). Focusing on the contact angle plot, we observe that the simulation predicts a gradual decay in the first three steps similar to that in the experiments. Further, similar to the experiments, we also observe that the relaxation time of the contact angles increases with every separation step. Quantitatively, there is a $250\%$ increase in the relaxation time between the first and the third step. The contact angle behavior in the last step does not agree with the experiments, with the simulations predicting a drastic increase in the angle. It is likely that simulations are predicting the correct trend, as practical challenges, including optical artefacts, limit the accuracy of the contact angle measurements in the terminal seconds leading to the complete separation of the bilayer.  Finally, the bilayer radius given by the simulation behaves similarly to those in the experiments (Fig.\ref{fig:Simulation}c). We also observe, similar to the experiments, the simulation predicts an increasing decay in the magnitude of the radius and a longer relaxation time of the bilayer with each step separation. 

The increase in the relaxation times of the contact angle and the radius, as well as the increasing decay magnitude of the bilayer across subsequent separation steps are intrinsic features of droplet interface bilayer separation under step strain. The rational for the increase in the relaxation time can be understood by combining Eqns. \ref{eq:bilayerForce} and \ref{eq:SeperationVelocity}, and expressing $\bar{F}_{ap}$ in terms of the initial bilayer radius ($R_{b,in}$) and the initial contact angle ($\theta_{b,in}$) at the start of the step. This gives,   
\begin{equation}
    \bar{v} \propto \bar{R}_{b,in}^2 - \bar{R}_{b,in}\sin{\theta_{b,in}} - \bar{R}_b^2 + \bar{R}_b\sin{\theta_{b,eq}}.
\end{equation}
Everything else remaining the same, the separation $\bar{v}$ decreases with initial radius, resulting in a longer decay times with each separation step. Similarly, the rationale for the increasing  magnitude of the radius decay ($\Delta R_b = R_{b,in}- R_{b,eq}$) can be identified by noting that $R_b$ reaches its equilibrium value of $R_{b,eq}$ when $v = 0$. Utilizing this fact and simplifying using binomial series expansion, we obtain the following expression,
\begin{align}\nonumber
\Delta \bar{R}_b &= \frac{1}{8\bar{R}_{b,in}} (\sin^2 \theta_{b,in} -\sin^2 \theta_{b,eq})\\ 
&- \frac{1}{2} (\sin \theta_{b,eq} -\sin \theta_{b,in}) + \mathcal{O}\left(\frac{\sin^3 \theta_b}{\bar{R}_{b,in}^2}\right)
 \end{align} 
 Clearly, $\Delta R_b$ scales inversely with $R_{b,in}$.

In addition to clarifying the different physical processes during bilayer separation, the simple mathematical model also serves the following purposes. Firstly, the model can be used to easily predict the separation behavior of different lipids in various conditions by modifying the equilibrium angles, the monolayer surface tension and the bulk viscosity.  Secondly, by modifying the force to rate relation, this model also gives a good framework to simulate the bilayer separation under a constant force or a constant rate. Thirdly, with simple modifications the governing equations can be adapted to study the bilayer separation between a pendant drop and a hard sphere (see \bit{Supplementary Materials}).

\section{Conclusions}\label{sec:conclusion}
In this manuscript we studied the separation of droplet interface bilayers under step strain using a custom build apparatus (I-DDiaSS). Initially we showed that the bilayer energy can be determined using principles of energy conservation in addition to the well established force balance method. Subsequently, we experimentally revealed the separation mechanics of the bilayer separation under step separation by tracking the evolution of the contact angle and the bilayer radius. Interestingly, the relaxation time of the contact angle and the bilayer radius, as well as the decay magnitude of the bilayer radius were observed to increase with each separation step. Through the analysis of the forces acting on the bilayer and the bilayer separation rates, we showed that the bilayer primarily separates through a peeling mechanism.  Finally, we also developed a mathematical model to successfully simulate the bilayer separation process. From the governing equations, we also showed that the separation velocity scales with the square of the initial bilayer radius and that the bilayer decay magnitude scales inversely with initial bilayer radius - the rationale for increasing relaxation times and bilayer decay magnitudes observed in the experiments. These  results improve our understanding of bilayer separation mechanics under step strain and  supplements the scientific efforts aimed at characterizing separation mechanics of bilayers under different separation modes \cite{frostad2014direct, brochard2003unbinding, chatkaew2009dynamics, bello2017lipid}.

There remain several opportunities for future work that would offer important extensions to the present work. First and foremost, it would be worthwhile to study the influence of bio-physically relevant molecules such as cholesterol on the separation mechanics of the lipid bilayers. Secondly, it is also important to investigate the effects of temperature on the reported separation mechanics. Finally, understanding the distribution of stress on the monolayer during the separation process, possibly using membrane tension probes such as FliptR\cite{colom2018fluorescent}, is also a promising direction for future research.

\bibliographystyle{vancouver}
\bibliography{References}

\begin{thebibliography}{10}

\bibitem{lombard2014once}
Lombard J.
\newblock Once upon a time the cell membranes: 175 years of cell boundary
  research.
\newblock Biology direct. 2014;9(1):1--35.

\bibitem{naumann2008protein}
Naumann RL, Knoll W.
\newblock Protein tethered lipid bilayer: An alternative mimic of the
  biological membrane (Mini Review).
\newblock Biointerphases. 2008;3(2):FA101--FA107.

\bibitem{bello2017lipid}
Bello J, Kim YR, Kim SM, Jeon TJ, Shim J.
\newblock Lipid bilayer membrane technologies: A review on single-molecule
  studies of DNA sequencing by using membrane nanopores.
\newblock Microchimica Acta. 2017;184(7):1883--1897.

\bibitem{gorter1925bimolecular}
Gorter E, Grendel F.
\newblock On bimolecular layers of lipoids on the chromocytes of the blood.
\newblock The Journal of experimental medicine. 1925;41(4):439.

\bibitem{singer1972fluid}
Singer SJ, Nicolson GL.
\newblock The fluid mosaic model of the structure of cell membranes.
\newblock Science. 1972;175(4023):720--731.

\bibitem{gambale1982properties}
Gambale F, Robello M, Usai C, Marchetti C.
\newblock Properties of ionic transport through phospholipid-glycolipid
  artificial bilayers.
\newblock Biochimica et Biophysica Acta (BBA)-Biomembranes.
  1982;693(1):165--172.

\bibitem{ter1993interaction}
ter BEEST MB, HOEKSTRA D.
\newblock Interaction of myelin basic protein with artificial membranes:
  parameters governing binding, aggregation and dissociation.
\newblock European journal of biochemistry. 1993;211(3):689--696.

\bibitem{villar2011formation}
Villar G, Heron AJ, Bayley H.
\newblock Formation of droplet networks that function in aqueous environments.
\newblock Nature nanotechnology. 2011;6(12):803--808.

\bibitem{barlow2018measuring}
Barlow NE, Kusumaatmaja H, Salehi-Reyhani A, Brooks N, Barter LM, Flemming AJ,
  et~al.
\newblock Measuring bilayer surface energy and curvature in asymmetric droplet
  interface bilayers.
\newblock Journal of the Royal Society Interface. 2018;15(148):20180610.

\bibitem{ohki1985divalent}
Ohki S, Ohshima H.
\newblock Divalent cation-induced phosphatidic acid membrane fusion. Effect of
  ion binding and membrane surface tension.
\newblock Biochimica et Biophysica Acta (BBA)-Biomembranes.
  1985;812(1):147--154.

\bibitem{lee2004lipids}
Lee AG.
\newblock How lipids affect the activities of integral membrane proteins.
\newblock Biochimica et Biophysica Acta (BBA)-Biomembranes.
  2004;1666(1-2):62--87.

\bibitem{lee2018revisiting}
Lee DW.
\newblock Revisiting the Interaction Force Measurement between Lipid Bilayers
  Using a Surface Forces Apparatus (SFA).
\newblock Journal of oleo science. 2018;67(11):1361--1372.

\bibitem{orozco2013interaction}
Orozco-Alcaraz R, Kuhl TL.
\newblock Interaction forces between DPPC bilayers on glass.
\newblock Langmuir. 2013;29(1):337--343.

\bibitem{beltramo2016millimeter}
Beltramo PJ, Van~Hooghten R, Vermant J.
\newblock Millimeter-area, free standing, phospholipid bilayers.
\newblock Soft Matter. 2016;12(19):4324--4331.

\bibitem{funakoshi2006lipid}
Funakoshi K, Suzuki H, Takeuchi S.
\newblock Lipid bilayer formation by contacting monolayers in a microfluidic
  device for membrane protein analysis.
\newblock Analytical chemistry. 2006;78(24):8169--8174.

\bibitem{evans1987physical}
Evans E, Needham D.
\newblock Physical properties of surfactant bilayer membranes: thermal
  transitions, elasticity, rigidity, cohesion and colloidal interactions.
\newblock Journal of Physical Chemistry. 1987;91(16):4219--4228.

\bibitem{bayley2008droplet}
Bayley H, Cronin B, Heron A, Holden MA, Hwang WL, Syeda R, et~al.
\newblock Droplet interface bilayers.
\newblock Molecular BioSystems. 2008;4(12):1191--1208.

\bibitem{thiam2012stability}
Thiam AR, Bremond N, Bibette J.
\newblock From stability to permeability of adhesive emulsion bilayers.
\newblock Langmuir. 2012;28(15):6291--6298.

\bibitem{alvarez2009non}
Alvarez NJ, Walker LM, Anna SL.
\newblock A non-gradient based algorithm for the determination of surface
  tension from a pendant drop: Application to low Bond number drop shapes.
\newblock Journal of colloid and interface science. 2009;333(2):557--562.

\bibitem{neeson2014compound}
Neeson MJ, Chan DY, Tabor RF.
\newblock Compound pendant drop tensiometry for interfacial tension measurement
  at zero bond number.
\newblock Langmuir. 2014;30(51):15388--15391.

\bibitem{najem2017mechanics}
Najem JS, Freeman EC, Yasmann A, Sukharev S, Leo DJ.
\newblock Mechanics of droplet interface bilayer “unzipping” defines the
  bandwidth for the mechanotransduction response of reconstituted MscL.
\newblock Advanced Materials Interfaces. 2017;4(3):1600805.

\bibitem{frostad2014direct}
Frostad JM, Seth M, Bernasek SM, Leal LG.
\newblock Direct measurement of interaction forces between charged
  multilamellar vesicles.
\newblock Soft matter. 2014;10(39):7769--7780.

\bibitem{hwang2007electrical}
Hwang WL, Holden MA, White S, Bayley H.
\newblock Electrical behavior of droplet interface bilayer networks:
  experimental analysis and modeling.
\newblock Journal of the American Chemical Society. 2007;129(38):11854--11864.

\bibitem{punnamaraju2011voltage}
Punnamaraju S, Steckl AJ.
\newblock Voltage control of droplet interface bilayer lipid membrane
  dimensions.
\newblock Langmuir. 2011;27(2):618--626.

\bibitem{lee2018static}
Lee Y, Lee HR, Kim K, Choi SQ.
\newblock Static and Dynamic Permeability Assay for Hydrophilic Small Molecules
  Using a Planar Droplet Interface Bilayer.
\newblock Analytical chemistry. 2018;90(3):1660--1667.

\bibitem{fleury2020enhanced}
Fleury JB.
\newblock Enhanced water permeability across a physiological droplet interface
  bilayer doped with fullerenes.
\newblock RSC Advances. 2020;10(33):19686--19692.

\bibitem{dixit2012droplet}
Dixit SS, Pincus A, Guo B, Faris GW.
\newblock Droplet shape analysis and permeability studies in droplet lipid
  bilayers.
\newblock Langmuir. 2012;28(19):7442--7451.

\bibitem{taylor2015direct}
Taylor GJ, Venkatesan GA, Collier CP, Sarles SA.
\newblock Direct in situ measurement of specific capacitance, monolayer
  tension, and bilayer tension in a droplet interface bilayer.
\newblock Soft matter. 2015;11(38):7592--7605.

\bibitem{venkatesan2015adsorption}
Venkatesan GA, Lee J, Farimani AB, Heiranian M, Collier CP, Aluru NR, et~al.
\newblock Adsorption kinetics dictate monolayer self-assembly for both lipid-in
  and lipid-out approaches to droplet interface bilayer formation.
\newblock Langmuir. 2015;31(47):12883--12893.

\bibitem{yanagisawa2013adhesive}
Yanagisawa M, Yoshida Ta, Furuta M, Nakata S, Tokita M.
\newblock Adhesive force between paired microdroplets coated with lipid
  monolayers.
\newblock Soft Matter. 2013;9(25):5891--5897.

\bibitem{chatkaew2009dynamics}
Chatkaew S, Georgelin M, Jaeger M, Leonetti M.
\newblock Dynamics of vesicle unbinding under axisymmetric flow.
\newblock Physical review letters. 2009;103(24):248103.

\bibitem{freeman2016mechanoelectrical}
Freeman E, Najem J, Sukharev S, Philen M, Leo D.
\newblock The mechanoelectrical response of droplet interface bilayer
  membranes.
\newblock Soft Matter. 2016;12(12):3021--3031.

\bibitem{holden2007functional}
Holden MA, Needham D, Bayley H.
\newblock Functional bionetworks from nanoliter water droplets.
\newblock Journal of the American Chemical Society. 2007;129(27):8650--8655.

\bibitem{leptihn2013constructing}
Leptihn S, Castell OK, Cronin B, Lee EH, Gross LC, Marshall DP, et~al.
\newblock Constructing droplet interface bilayers from the contact of aqueous
  droplets in oil.
\newblock Nature protocols. 2013;8(6):1048--1057.

\bibitem{poulos2010automatable}
Poulos J, Portonovo S, Bang H, Schmidt J.
\newblock Automatable lipid bilayer formation and ion channel measurement using
  sessile droplets.
\newblock Journal of Physics: Condensed Matter. 2010;22(45):454105.

\bibitem{bochner2017droplet}
Bochner~de Araujo S, Merola M, Vlassopoulos D, Fuller GG.
\newblock Droplet coalescence and spontaneous emulsification in the presence of
  asphaltene adsorption.
\newblock Langmuir. 2017;33(40):10501--10510.

\bibitem{brochard2003unbinding}
Brochard-Wyart F, de~Gennes PG.
\newblock Unbinding of adhesive vesicles.
\newblock Comptes Rendus Physique. 2003;4(2):281--287.

\bibitem{waugh1982surface}
Waugh RE.
\newblock Surface viscosity measurements from large bilayer vesicle tether
  formation. II. Experiments.
\newblock Biophysical journal. 1982;38(1):29.

\bibitem{colom2018fluorescent}
Colom A, Derivery E, Soleimanpour S, Tomba C, Dal~Molin M, Sakai N, et~al.
\newblock A fluorescent membrane tension probe.
\newblock Nature chemistry. 2018;10(11):1118--1125.

\end{thebibliography}

\end{document}